\documentclass[%
10pt,
aps,
prl,
twocolumn,
superscriptaddress,
showpacs,
altaffilletter,
endnotes
nofootinbib,
 amsmath,
 amssymb,
]{revtex4-2}

\usepackage{graphicx}
\usepackage{dcolumn}
\usepackage{bm}
\usepackage{hyperref}
\hypersetup{colorlinks=true, citecolor=blue, urlcolor=blue, linkcolor=blue}

\usepackage[alsoload=hep]{siunitx} 

\usepackage{xcolor} 
\usepackage{comment} 
\usepackage{booktabs}

\usepackage[printwatermark]{xwatermark}
\usepackage{bm} 
\usepackage[ampersand]{easylist} 

\usepackage{amsmath}
\usepackage{xspace}

\begin{document}

\newcommand{\gm}{\ensuremath{(g-2)}\xspace}
\newcommand{\wa}{\ensuremath{\omega_{a}}\xspace}
\newcommand{\wam}{\ensuremath{\omega_{a}^{m}}\xspace}
\renewcommand{\wp}{\ensuremath{\omega_{p}}\xspace}
\newcommand{\opprime}{\ensuremath{\omega'^{}_p}\xspace}
\newcommand{\opprimetilde}{\ensuremath{\tilde{\omega}'^{}_p}\xspace}
\newcommand{\optilde}{\ensuremath{\tilde{\omega}^{}_p}\xspace}
\renewcommand{\amu}{\ensuremath{a_{\mu}}\xspace}
\newcommand{\Rmu}{\ensuremath{{\mathcal R}_\mu}\xspace}
\newcommand{\Rmuprime}{\ensuremath{{\mathcal R}'^{}_\mu}\xspace}
\newcommand{\gmtwo}{\ensuremath{g\!-\!2}\xspace}
\newcommand{\rmagic}{\ensuremath{r_{0}}\xspace}
\newcommand{\cmagic}{\ensuremath{c_{0}}\xspace}
\newcommand{\pmagic}{\ensuremath{p_{0}}\xspace}
\newcommand{\gammamagic}{\ensuremath{\gamma_{0}}\xspace}
\newcommand{\betamagic}{\ensuremath{\beta_{0}}\xspace}
\newcommand{\dipbfield}{\ensuremath{B_{0}}\xspace}
\newcommand{\runone}{Run-1\xspace}
\newcommand{\runtwo}{Run-2\xspace}
\newcommand{\runthree}{Run-3\xspace}
\newcommand{\runfour}{Run-4\xspace}
\newcommand\runonea{Run-1a\xspace}
\newcommand\runoneb{Run-1b\xspace}
\newcommand\runonec{Run-1c\xspace}
\newcommand\runoned{Run-1d\xspace}

\newcommand{\precession}{E989wapaper}
\newcommand{\field}{E989fieldpaper}
\newcommand{\BD}{E989SRBDpaper}

\title{Measurement of the Positive Muon Anomalous Magnetic Moment to 0.46\,ppm}

\affiliation{Argonne National Laboratory, Lemont, IL, USA}
\affiliation{Boston University, Boston, MA, USA}
\affiliation{Brookhaven National Laboratory, Upton, NY, USA}
\affiliation{Budker Institute of Nuclear Physics, Novosibirsk, Russia}
\affiliation{Center for Axion and Precision Physics (CAPP) / Institute for Basic Science (IBS), Daejeon, Republic of Korea}
\affiliation{Cornell University, Ithaca, NY, USA}
\affiliation{Fermi National Accelerator Laboratory, Batavia, IL, USA}
\affiliation{INFN Gruppo Collegato di Udine, Sezione di Trieste, Udine, Italy}
\affiliation{INFN, Laboratori Nazionali di Frascati, Frascati, Italy}
\affiliation{INFN, Sezione di Napoli, Napoli, Italy}
\affiliation{INFN, Sezione di Pisa, Pisa, Italy}
\affiliation{INFN, Sezione di Roma Tor Vergata, Roma, Italy}
\affiliation{INFN, Sezione di Trieste, Trieste, Italy}
\affiliation{Istituto Nazionale di Ottica - Consiglio Nazionale delle Ricerche, Pisa, Italy}
\affiliation{Department of Physics and Astronomy, James Madison University, Harrisonburg, VA, USA}
\affiliation{Institute of Physics and Cluster of Excellence PRISMA+, Johannes Gutenberg University Mainz, Mainz, Germany}
\affiliation{Joint Institute for Nuclear Research, Dubna, Russia}
\affiliation{Department of Physics, Korea Advanced Institute of Science and Technology (KAIST), Daejeon, Republic of Korea}
\affiliation{Lancaster University, Lancaster, United Kingdom}
\affiliation{Michigan State University, East Lansing, MI, USA}
\affiliation{North Central College, Naperville, IL, USA}
\affiliation{Northern Illinois University, DeKalb, IL, USA}
\affiliation{Northwestern University, Evanston, IL, USA}
\affiliation{Regis University, Denver, CO, USA}
\affiliation{Scuola Normale Superiore, Pisa, Italy}
\affiliation{School of Physics and Astronomy, Shanghai Jiao Tong University, Shanghai, China}
\affiliation{Tsung-Dao Lee Institute, Shanghai Jiao Tong University, Shanghai, China}
\affiliation{Institut für Kern - und Teilchenphysik, Technische Universit\"at Dresden, Dresden, Germany}
\affiliation{Universit\`a del Molise, Campobasso, Italy}
\affiliation{Universit\`a di Cassino e del Lazio Meridionale, Cassino, Italy}
\affiliation{Universit\`a di Napoli, Napoli, Italy}
\affiliation{Universit\`a di Pisa, Pisa, Italy}
\affiliation{Universit\`a di Roma Tor Vergata, Rome, Italy}
\affiliation{Universit\`a di Trieste, Trieste, Italy}
\affiliation{Universit\`a di Udine, Udine, Italy}
\affiliation{Department of Physics and Astronomy, University College London, London, United Kingdom}
\affiliation{University of Illinois at Urbana-Champaign, Urbana, IL, USA}
\affiliation{University of Kentucky, Lexington, KY, USA}
\affiliation{University of Liverpool, Liverpool, United Kingdom}
\affiliation{Department of Physics and Astronomy, University of Manchester, Manchester, United Kingdom}
\affiliation{Department of Physics, University of Massachusetts, Amherst, MA, USA}
\affiliation{University of Michigan, Ann Arbor, MI, USA}
\affiliation{University of Mississippi, University, MS, USA}
\affiliation{University of Oxford, Oxford, United Kingdom}
\affiliation{University of Rijeka, Rijeka, Croatia}
\affiliation{Department of Physics, University of Texas at Austin, Austin, TX, USA}
\affiliation{University of Virginia, Charlottesville, VA, USA}
\affiliation{University of Washington, Seattle, WA, USA}
\author{B.~Abi}  \affiliation{University of Oxford, Oxford, United Kingdom}
\author{T.~Albahri}  \affiliation{University of Liverpool, Liverpool, United Kingdom}
\author{S.~Al-Kilani}  \affiliation{Department of Physics and Astronomy, University College London, London, United Kingdom}
\author{D.~Allspach}  \affiliation{Fermi National Accelerator Laboratory, Batavia, IL, USA}
\author{L.~P.~Alonzi}  \affiliation{University of Washington, Seattle, WA, USA}
\author{A.~Anastasi} \thanks{Deceased} \affiliation{INFN, Sezione di Pisa, Pisa, Italy}
\author{A.~Anisenkov} \altaffiliation[Also at ]{Novosibirsk State University}  \affiliation{Budker Institute of Nuclear Physics, Novosibirsk, Russia}
\author{F.~Azfar}  \affiliation{University of Oxford, Oxford, United Kingdom}
\author{K.~Badgley}  \affiliation{Fermi National Accelerator Laboratory, Batavia, IL, USA}
\author{S.~Bae{\ss}ler} \altaffiliation[Also at ]{Oak Ridge National Laboratory}  \affiliation{University of Virginia, Charlottesville, VA, USA}
% ...
\collaboration{The Muon \gmtwo Collaboration} \noaffiliation
\vskip 0.25cm

\date{\today}

\begin{abstract}
\medskip
We present the first results of the Fermilab Muon \gmtwo Experiment for the positive muon magnetic
anomaly $a_\mu \equiv (g_\mu-2)/2$.  The anomaly is determined from the precision measurements of two angular frequencies.
Intensity variation of high-energy positrons from muon decays directly encodes the difference frequency
$\omega_a$ between the spin-precession and cyclotron frequencies for polarized muons in a magnetic storage ring.
The storage ring magnetic field is measured using nuclear magnetic resonance probes calibrated in terms of the equivalent proton spin precession frequency \opprimetilde in a spherical water sample at 34.7$^{\circ}$C.
The ratio $\wa / \opprimetilde$, together with known fundamental constants, determines $a_\mu({\rm FNAL}) = 116\,592\,040(54)\times  10^{-11}$ (0.46\,ppm).  The result
is 3.3 standard deviations greater than the standard model prediction and is in excellent agreement with the previous
Brookhaven National Laboratory (BNL) E821 measurement.  After combination with previous measurements of both $\mu^+$ and $\mu^-$, the new experimental average of $a_\mu({\rm Exp}) = 116\,592\,061(41)\times  10^{-11}$ (0.35\,ppm) increases
the tension between experiment and theory to 4.2 standard deviations.

\end{abstract}

\maketitle 

\section{\label{sec:intro} Introduction}

The magnetic moments of the electron and muon
\begin{equation*}
  \vec \mu_{\ell} = g_{\ell} \left(\frac{q}{2m_{\ell}}\right) \vec s\,
  \ \ {\rm where} \ \  g_{\ell} =
2(1+a_{\ell}),
  \label{eq:MDM}
\end{equation*}
($\ell = e,\mu$) have played
an important role
in the development of the standard model (SM).
One of the triumphs of the Dirac
equation~\cite{Dirac:1928ej}
was its prediction for the electron that $g_e =2$.  Motivated in part by anomalies in the hyperfine structure
of hydrogen~\cite{Nagel:1947aa,Nafe:1947zz}, Schwinger~\cite{Schwinger:1948iu}
proposed an additional contribution to the electron magnetic moment from
a radiative correction, predicting the anomaly~\footnote{The scalar
quantity $a_{\ell}$ is the
magnetic anomaly, but is also commonly referred to as the ``anomaly'' or
the ``anomalous magnetic moment''
in the literature.}
$a_e = \alpha/2\pi \simeq 0.00116$ in agreement with experiment~\cite{Kusch:1948aa}.

The first muon spin
rotation experiment that observed parity violation in muon
decay~\cite{Garwin:1957hc} determined
that, to within 10\%, $g_\mu = 2$, which was subsequently measured with higher precision~\cite{0370-1298-70-7-412}. A more precise experiment~\cite{Garwin:1960zz} confirmed Schwinger's prediction for the muon  anomaly and thereby established for the first time the notion that a muon behaved like a heavy electron in a magnetic field.
This evidence, combined with
the discovery of the muon neutrino~\cite{Danby:1962nd},
pointed to the generational structure of the SM.

\begin{figure}[t]
\centering
\includegraphics[width=\columnwidth]{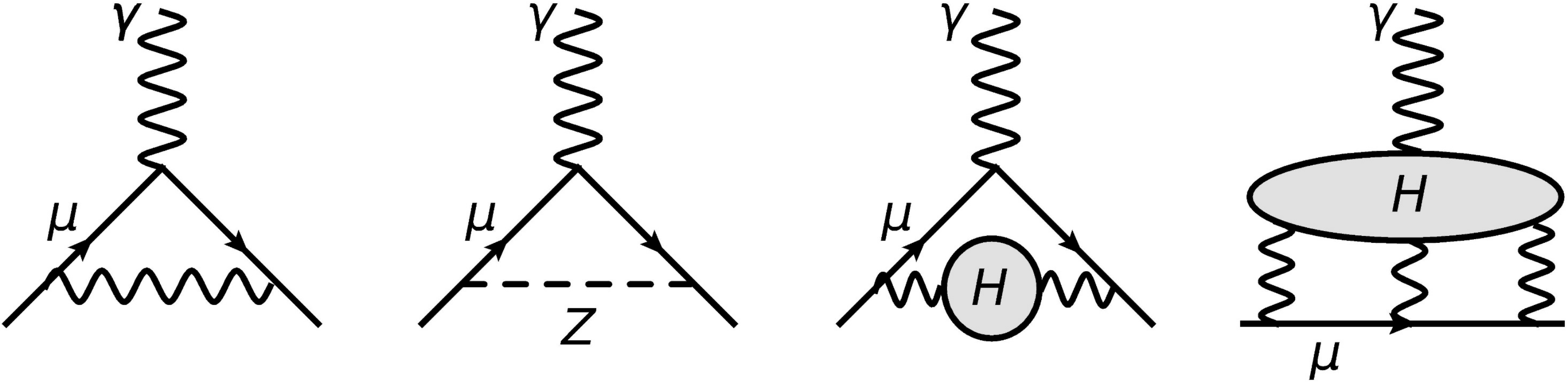}
\caption{Feynman diagrams of representative SM contributions to the muon anomaly. From left to right: first-order QED and weak processes, leading-order hadronic (H) vacuum polarization and hadronic light-by-light contributions.}\label{fig:feynman}
\end{figure}

The  SM contributions to the muon anomaly, as illustrated in Fig.~\ref{fig:feynman}, include electromagnetic, strong, and weak interactions that arise from virtual effects
involving photons,  leptons, hadrons, and the $W$, $Z$, and Higgs bosons~\cite{Jegerlehner:2017gek}.
Recently, the international theory community published
a comprehensive~\footnote{The value is based on evaluating hadronic vacuum polarization contributions via $e^+e^-\rightarrow \text{hadrons}$ data.
Lattice QCD calculations of the hadronic vacuum polarization show promising improvements~\cite{Aoyama:2020ynm,Chakraborty:2017tqp,Borsanyi:2017zdw,Blum:2018mom,Giusti:2019xct,Shintani:2019wai,Davies:2019efs,Gerardin:2019rua,Aubin:2019usy,Giusti:2019hkz,Borsanyi:2020mff,Lehner:2020crt}. The lattice world average determined in Ref.~\cite{Aoyama:2020ynm} is consistent with the data-driven result used for the number in the main text, but has a higher central value and larger uncertainty. Further scrutiny and improvements of lattice results are expected.}
SM prediction~\cite{Aoyama:2020ynm} for the muon anomaly, finding
$a_\mu(\text{SM}) = 116\,591\,810(43) \times 10^{-11}$ (0.37\,ppm).
It is based on state-of-the-art evaluations of the contributions from quantum electrodynamics (QED) to tenth order~\cite{Aoyama:2012wk,Aoyama:2019ryr}, hadronic vacuum polarization
\cite{Davier:2017zfy,Keshavarzi:2018mgv,Colangelo:2018mtw,Hoferichter:2019gzf,Davier:2019can,Keshavarzi:2019abf,Kurz:2014wya},
hadronic light-by-light
\cite{Melnikov:2003xd,Masjuan:2017tvw,Colangelo:2017fiz,Hoferichter:2018kwz,Gerardin:2019vio,Bijnens:2019ghy,Colangelo:2019uex,Pauk:2014rta,Danilkin:2016hnh,Jegerlehner:2017gek,Knecht:2018sci,Eichmann:2019bqf,Roig:2019reh,Blum:2019ugy,Colangelo:2014qya},
and electroweak
processes~\cite{Jackiw:1972jz,Bars:1972pe,Fujikawa:1972fe,Czarnecki:2002nt,Gnendiger:2013pva}.

The measurement of \amu\ has become increasingly precise through a series of
innovations employed by three experimental campaigns at
CERN~\cite{Charpak:1962zz,Bailey:1969pr,Bailey:1978mn}
and more recently at Brookhaven (BNL E821)~\cite{Bennett:2006fi}.
The BNL net result, with its 0.54\,ppm precision, is larger than $a_\mu(\text{SM})$
by 3.7 standard deviations ($\sigma$).
While the electron magnetic anomaly has been measured to fractions of a part
per billion~\cite{Hanneke:2008tm},
the relative contribution of virtual heavy particles in many cases scales as $(m_\mu/m_e)^2\simeq 43,000$. This is the case e.g.~for the $W$ and $Z$ bosons of the SM and many hypothetical new particles, and it gives the muon anomaly a significant advantage when searching for effects of new heavy physics.
Because the BNL result
hints at  physics not included in the SM,
Experiment E989~\cite{Grange:2015fou} at Fermilab was constructed
to independently confirm or refute that finding.
In this paper, we report our first result with a precision of 0.46\,ppm.
Separate papers provide analysis details on the muon precession~\cite{\precession}, the beam dynamics corrections~\cite{\BD}, and the magnetic field~\cite{\field} determination.

\section{\label{sec:bd_overview} Experimental Method}
The experiment follows the BNL concept~\cite{Bennett:2006fi} and uses the same 1.45\,T superconducting storage ring (SR) magnet~\cite{Danby:2001eh}, but it benefits from  substantial improvements. These include a 2.5 times improved magnetic field intrinsic uniformity, detailed beam storage simulations, and state-of-the-art tracking, calorimetry, and field metrology for the measurement of the beam properties,  precession frequency, and  magnetic field~\cite{Grange:2015fou}.

The Fermilab Muon Campus delivers 16 highly polarized, 3.1\,GeV/$c$, ${\sim}120$\,ns long positive muon beam bunches every 1.4\,s into the SR.  A fast pulsed-kicker magnet deflects the muon bunch into a  9-cm-diameter storage aperture, resulting in $\approx 5000$ stored muons per fill.
The central orbit has a radius of $R_0 =7.112$\,m and the cyclotron period is 149.2\,ns.
Four sections of electrostatic quadrupole (ESQ) plates provide weak focusing for vertical confinement.

The muon spins precess in the magnetic field at a rate greater than the cyclotron frequency.  The anomalous precession frequency~\footnote{The rate of
change of the angle between spin and
momentum vectors
is equal to $\vec\omega_a$ only if
$\vec\omega_s$ and $\vec \omega_c$ are parallel.
The angle between $\vec\omega_s$ and $\vec\omega_c$ is always small, and the
rate of oscillation of $\vec\beta$ out of pure circular motion is fast compared to \wa.}
in the presence of the electric $\vec{E}$ and magnetic $\vec{B}$ fields
of the SR is
\begin{equation}
\begin{aligned}
\vec{\omega}_{a}\equiv\vec{\omega}_{s}-\vec{\omega}_{c} &=-\frac{q}{m_{\mu}}\Bigg[ a_{\mu} \vec{B}-a_{\mu}\left(\frac{\gamma}{\gamma+1}\right)(\vec{\beta} \cdot \vec{B}) \vec{\beta} \\
& \quad \left.-\left(a_{\mu}-\frac{1}{\gamma^{2}-1}\right) \frac{\vec{\beta} \times \vec{E}}{c} \right].
\label{eq:omega}
\end{aligned}
\end{equation}
For horizontally circulating muons in a vertical magnetic field, $\vec{\beta}\cdot\vec{B} = 0$; this condition is approximately met in our SR.
At the muon central momentum $p_0$, set such that  $\gamma_\mu = \sqrt{(1+1/a_\mu)} \approx 29.3$, the third term vanishes.

In-vacuum straw tracker stations located at azimuthal angle $\phi = 180^{\circ}$ and $270^{\circ}$ with respect to the injection point provide nondestructive, time-in-fill dependent beam profiles  $M(x,y,\phi,t)$ by extrapolation of decay positron trajectories to their upstream radial tangency points within the storage aperture~\footnote{The coordinate system is with respect to the center of the storage volume at radius $R_0$, with $x$ radially outward, $y$ vertically up, and $\phi$ increasing clockwise when viewed from above.}.  These profiles determine the betatron oscillation parameters necessary for beam dynamics corrections and the precession data fits discussed below.

Twenty-four calorimeters~\cite{Fienberg:14,Khaw:2019yzq,Kaspar:2016ofv}, each containing a $9\times 6$ array of PbF$_{2}$ crystals, detect the inward-spiraling decay positrons.  When a signal in a silicon photomultiplier (SiPM) viewing any crystal exceeds $\sim 50$~MeV, the data-acquisition system stores the 54 waveforms from that calorimeter in a set time window around the event.
Decay positron hit times and energies are derived from reconstruction of the waveforms.

The magnetic field is measured using pulsed proton NMR, calibrated in terms of the equivalent precession frequency $\omega_{p}’(T_{r})$ of a proton shielded in a spherical sample of water at a reference temperature $T_{r}=34.7^{\circ}$C.  The magnetic field $B$ is determined from the precession frequency and shielded proton magnetic moment, $\mu_{p}’(T_{r})$ using $\hbar\omega_{p}’=2\mu_{p}’B$. The
muon anomaly can then be obtained
 from~\footnote{We use the shielded proton-to-electron magnetic moment ratio~\cite{Phillips:1977} and the electron $g$-factor~\cite{Hanneke:2010au} measurement. The CODATA-2018 result is used for the muon-to-electron mass ratio~\cite{CODATA:2018}, which is determined from bound-state QED theory and measurements described in~\cite{Liu:1999iz}. The QED factor $\mu^{}_e(H)/\mu_e$ is computed by theory with negligible uncertainty~\cite{CODATA:2018}.}
\begin{equation}
\begin{aligned}
                a^{}_{\mu} = \frac{\omega^{}_a}{\opprimetilde(T_{r})} \frac{\mu'^{}_p(T_{r})}{\mu^{}_e(H)} \frac{{\mu^{\
}_e(H)}}{\mu^{}_e} \frac{m^{}_{\mu}}{m^{}_e} \frac{g^{}_e}{2},
\label{eq:amueq}
\end{aligned}
\end{equation}
where our collaboration measures the two quantities to form the ratio
\begin{equation}
\Rmu^{'} \equiv \frac{\wa}{\opprimetilde(T_{r})}.
\label{eq:mathcalR}
\end{equation}

The \runone data, collected in 2018, are grouped into four subsets (a -- d) that are distinguished by unique kicker and  ESQ voltage  combinations.
The ratio $\Rmu'$ can be conceptually written in terms of measured quantities and corrections as
\begin{equation}
\Rmu^{'}  \approx \frac{f_{\rm clock}~  \wam  \left (1 + C_e + C_p + C_{ml} + C_{pa}  \right )}{f_{\rm calib}~\langle \omega_{p}’(x,y,\phi)\times M(x,y,\phi)\rangle (1+B_{k}+B_{q}) }.
\label{eq:Rcomponents}
\end{equation}
The numerator includes the master clock unblinding factor $f_{\text{clock}}$, the measured precession frequency \wam,  and four beam-dynamics corrections, $C_i$.  We  deconstruct $\opprimetilde(T)$ into the absolute NMR calibration procedure (indicated by $f_{\rm calib}$) and the field maps, which are weighted by the detected positrons and the muon distribution averaged over several timescales ($\langle \omega_{p}’(x,y,\phi)\times M(x,y,\phi)\rangle$).  The result must be  corrected for two  fast magnetic transients $B_i$ that are synchronized to the injection.

Damage to two of the 32 ESQ high-voltage resistors was discovered after completion of \runone.  This led to slower-than-designed charging of one of the quadrupole sections, spoiling the symmetry of the electric field early in each fill.
The impact of this is accounted for in the analysis presented.
Brief summaries of the terms in Eq.~\ref{eq:Rcomponents} follow.

\section{\label{sec:precession} Anomalous precession frequency}

$\bm{f_{\text{clock}}}$:   A single 10\,MHz, GPS-disciplined master clock drives both the \wa and \opprimetilde measurements. The clock has a one-week Allan deviation~\cite{AllanVarRef}  of 1\,ppt.
Two frequencies derived from this clock provide the 61.74\,MHz field reference and a blinded ``$(40-\epsilon)$\,MHz'' used for the \wa precession measurement.
A blinding factor in the range $\pm25$\,ppm was set and monitored by individuals external to our collaboration. $f_{\text{clock}}$ is the unblinding conversion factor; its uncertainty is negligible.

$\bm{\omega_a^m}$:
The signature of muon spin precession stems from parity violation in $\mu^{+}$ decay, which correlates the muon spin and the positron emission directions in the $\mu^{+}$ rest frame.  When boosted to the lab frame, this correlation  modulates the $e^{+}$ energy ($E$) spectrum at the relative precession frequency \wa between the muon spin and momentum directions.
The rate of detected positrons with $E>E_{\rm th}$ as a function of time $t$ into the muon fill then varies as
\begin{multline}
N(t)=N_0\eta_{N}(t) e^{-t / \gamma\tau_\mu} \\
\times\left[1+A \eta_{A}(t)\cos\left(\omega_{a}t + \varphi_0 + \eta_{\phi}(t)\right)\right],
	\label{eq:wiggle_func}
\end{multline}
where $\gamma\tau_\mu$ is the time-dilated muon lifetime ($\approx64.4\,\mu$s), $N_0$ is the normalization, $A$ is the average weak-decay asymmetry, and $\varphi_0$ is the ensemble average phase angle at injection.
The latter three parameters all depend on $E_{\rm th}$.  The $\eta_{i}$ terms model effects from
betatron oscillations of the beam, and are not required in their absence.  This beam motion couples with detector acceptance to modulate the rate and the average energy, and hence the average asymmetry and phase, at specific frequencies.  The coherent betatron oscillation (CBO) in the radial direction dominates the modulation.

The CBO, aliased vertical width (VW), and vertical mean ($\langle y \rangle$) frequencies  are well measured, and the $\eta_{i}$ terms are well modeled and minimally correlated in fits for \wa.

An accurate fit to the data also  requires accounting for the continuous loss of muons over a fill, also weakly coupled to \wa.  Coincident minimum-ionizing energies in three sequential calorimeters provide a signal to determine the time dependence of muon losses.

Two complementary reconstruction algorithms transform the digitized SiPM waveforms into positron energies and arrival times. In the  ``local'' approach, waveforms are template-fit to identify all pulses in each crystal, which are then clustered based on a time window.  In the ``global'' approach, waveforms in a $3\times 3$ array of crystals centered on a local maximum in time and position are template-fit simultaneously.  After subtraction of the fit from the waveforms, that algorithm iterates to test for any missed pulses from multiparticle pileup.
To avoid biasing \wa, we stabilize the calorimeter energy measurement within a muon fill by correcting the energy reconstruction algorithm on the SiPM pixel recovery timescale (up to tens of nanoseconds) and the fill timescale (700\,$\mu$s) using a laser-based monitoring system~\cite{Anastasi:2019lxf}.  The system also provides long-term (many-days) gain corrections.
The two reconstructed positron samples are used in four independent extractions of \wa in which each $e^{+}$ contribution to the time series is weighted by its energy-dependent asymmetry; this is the optimal approach~\cite{Bennett:2007zzb}. Seven other determinations using additional methods agree well~\cite{\precession}.
Each time series is modified to statistically correct for contributions of unresolved pileup clusters that result from multiple positrons proximate in space and time.
The analyses employ one of three data-driven techniques to correct for pileup, which would otherwise bias \wa.

\begin{figure}[tb]
\centering
\includegraphics[width=\columnwidth]{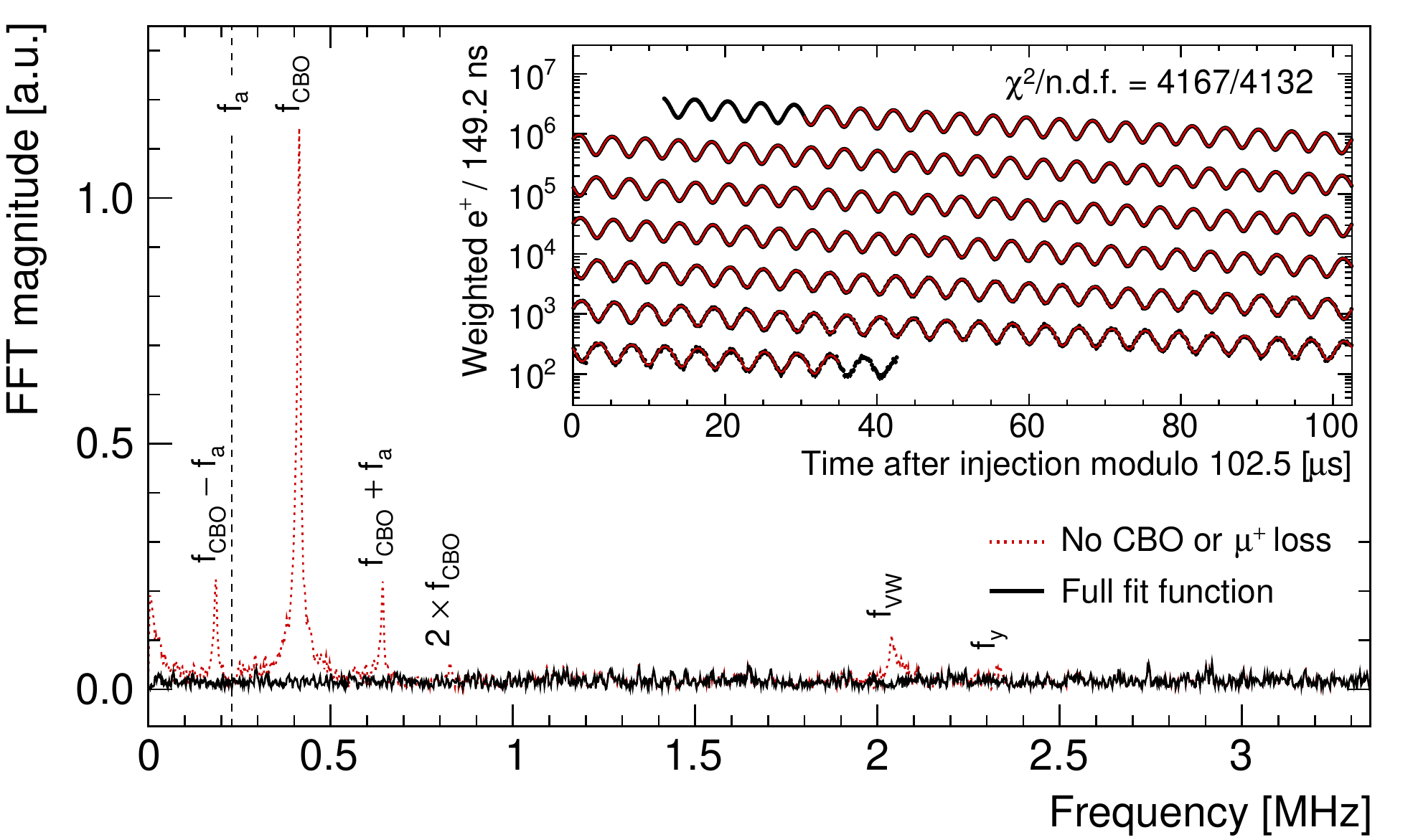}
\caption{Fourier transform of the residuals from a  time-series fit following Eq.~\ref{eq:wiggle_func} but neglecting betatron motion and muon loss (red dashed); and from the full fit (black).  The peaks correspond to the neglected betatron frequencies and  muon loss. Inset: Asymmetry weighted $e^{+}$ time spectrum (black) from the  \runonec run group fit with the full fit function (red) overlaid.}
\label{fig:precession}
\end{figure}

A $\chi^{2}$ minimization of the data model of Eq.~\ref{eq:wiggle_func} to the reconstructed time series determines the measured ($m$) quantity \wam.
The model fits the data well (see inset to Fig.~\ref{fig:precession}), producing reduced $\chi^{2}$s consistent with unity. Fourier transforms of the fit residuals show no unmodeled frequency components, see Fig.~\ref{fig:precession}.  Without the $\eta_i$ terms and the muon loss function in the model, strong signals emerge in the residuals at expected frequencies.

The dominant systematic uncertainties on \wa arise from  uncertainties in the pileup and gain correction factors, the modeling of the functional form of the CBO decoherence, and in the $\omega_{\text{CBO}}(t)$ model.  Scans varying the fit start and  stop times and across individual calorimeter stations showed no significant variation in any of the four run groups~\cite{\precession}.

The measured frequency \wam requires
four corrections, $C_i$, for interpretation as the anomalous precession frequency \wa of Eq.~\ref{eq:amueq}.
The details are found in Ref.~\cite{\BD}.

$\bm{C_e}$:
The electric-field correction $C_e$ from the last term in Eq.~\ref{eq:omega} depends on the distribution of equilibrium radii $x_e = x-R_{0}$, which translates to the muon beam momentum distribution via $\Delta p/p_0 \cong x_e(1-n)/R_0$, where $n$ is the field index determined by the ESQ  voltage~\cite{\BD}.  A Fourier analysis~\cite{Orlov:2002ag,\BD} of the decoherence rate of the incoming bunched beam as measured by the calorimeters provides the momentum distribution and determines the mean equilibrium radius $\langle x_e \rangle \approx 6$\,mm and the width $\sigma_{x_e} \approx 9$\,mm.  The final correction factor is
$C_e = 2n(1-n)\beta^2  \langle x^2_e\rangle / R^2_0$, where
$\langle x_e^2 \rangle = \sigma_{x_e}^2 + \langle x_e \rangle^2$.

$\bm{C_p}$:  A pitch correction $C_p$ is required to account for the vertical betatron oscillations that lead to a nonzero average value of the  $\vec{\beta}\cdot\vec{B}$ term in Eq.~\ref{eq:omega}.
The expression $C_p =   n \langle A_y^2\rangle / 4R_0^2$ determines the pitch correction factor~\cite{Farley:1973sc,\BD}. The acceptance-corrected vertical amplitude $A_y$ distribution in the above expression is measured by the trackers.

Extensive simulations determined the uncertainties $\delta C_e$ and $\delta C_p$ arising from the
geometry and alignment of the plates, as well as their voltage uncertainties and nonlinearities.
The nonuniform kicker time profile applied to the finite-length incoming muon bunch results in a correlation introducing the largest uncertainty on  $C_e$.

$\bm{C_{ml}}$:
Any bias in the average phase of muons that are lost compared to those that remain stored creates a time dependence to the phase factor $\varphi_0$ in Eq.~\ref{eq:wiggle_func}. Beamline simulations predict a phase-momentum correlation $d\varphi_0/dp = (-10.0 \pm 1.6)$\,mrad$/(\%\Delta p/p_0)$ and losses are known to be momentum dependent.
We verified the correlation by fitting  precession data from short runs in which the storage ring magnetic field, and thus the central stored momentum $p_0$, varied by $\pm0.67\%$ compared to its nominal setting.
Next, we measured the relative rates of muon loss ($ml$) versus momentum in dedicated runs in which muon distributions were heavily biased toward high or low momenta using upstream collimators.
Coupling the measured rate of muon loss in \runone to these two correlation factors determines the correction factor $C_{ml}$.

$\bm{C_{pa}}$:
The phase term $\varphi_0$ in Eq.~\ref{eq:wiggle_func}  depends on the muon decay coordinate $(x,y,\phi)$  and positron energy, but the precession frequency \wa does not.  If the stored muon average transverse distribution and the detector gains are stable throughout a fill, that average phase remains constant.
The two damaged resistors in the ESQ system caused slow changes to the  muon distribution during the first $\sim 100\,\mu$s of the measuring period.
An extensive study of this effect involved:
a) generation of  phase, asymmetry, and acceptance maps for each calorimeter as a function of muon decay coordinate and positron energy from simulations utilizing our {\tt GEANT}-based model of the ring ({\tt gm2ringsim});
b) extraction of the time dependence of the optical lattice around the ring from the {\tt COSY} simulation package and {\tt gm2ringsim};
c) folding the azimuthal beam distribution derived from tracker and optics simulations with the phase, asymmetry, and acceptance maps to determine a net effective phase shift versus time-in-fill, $\varphi_0(t)$; and
d) application of this time-dependent phase shift to precession data fits to determine the phase-acceptance ($pa$) correction $C_{pa}$.
The use of multiple approaches confirmed the conclusions; for details, see Ref.~\cite{\BD}.  The damaged resistors were replaced after \runone, which significantly reduces the dominant contribution to $C_{pa}$ and the overall magnitude of muon losses.

\section{Magnetic field determination}
A suite of pulsed-proton NMR probes, each optimized for a different function in the analysis chain, measures the magnetic field strength~\cite{\field}.
Every $\sim$3 days  during data taking, a 17-probe NMR trolley~\cite{Corrodi:2020sav} measures the field at about 9000 locations in azimuth to provide a set of 2D field maps.
378 pulsed-NMR probes, located 7.7\,cm above and below the storage volume, continuously monitor the field at 72 azimuthal positions, called stations.
The trolley and fixed probes use petroleum jelly as an NMR sample.
The probe signals are digitized and analyzed~\cite{Ran:2021freq} to extract a precession frequency proportional to the average magnetic field over the NMR sample volume.
A subset of probes is used to provide feedback to the magnet power supply to stabilize the field.

{\bf Calibration procedure} $\bm{f_{\text{calib}}}$:
The primary calibration uses a probe with a cylindrical water sample. Corrections are required to relate its frequencies to the precession frequency expected from a proton in water at the reference temperature 34.7$^{\circ}$C. Studies of the calibration probe in an MRI solenoid precisely determine corrections for sample shape, temperature, and magnetization of probe materials to an uncertainty of 15\,ppb.
Cross-calibrations to an absolute $^{3}$He magnetometer~\cite{Farooq:2020swf} confirm the corrections to better than 38\,ppb.

The calibration probe is installed on a translation stage in the SR vacuum. We repeatedly swap the calibration
probe and a trolley probe into the same location, compensating for changes of the SR field. This procedure determines calibration offsets between individual trolley probes and the equivalent $\omega_{p}'$ values.
The offsets are due primarily to differences in diamagnetic shielding of protons in water versus petroleum jelly, sample shape, and
magnetic perturbations from magnetization of the materials used in the probes and trolley body.
The trolley probe calibration offsets are determined with an average uncertainty of 30\,ppb.

{\bf Field Tracking} ($\omega_p^\prime(x,y,\phi)$):
The 14 \runone trolley field measurements bracket muon storage intervals $t_k$ to $t_{k+1}$.
They provide a suite of 2D multipole moments (dipole, normal quadrupole, skew quadrupole, ...), which the fixed probes track.
The fixed probes provide five independent moments (four moments for some stations) that track the field over 5$^\circ$ in azimuth for each station.
The trolley moments are interpolated for times between the trolley runs, and the fixed probes continuously track changes to five  lower-order moments~\cite{\field}.
The fixed probe and trolley measurements are synchronized when the trolley passes, averaged over  each 5$^{\circ}$ azimuthal segment.  The trolley run at time $t_{k+1}$ yields a second set of moments $m^{\text{tr}}_{i}(t_{k+1})$.
The fixed probe moments    $m^{\text{fp}}_{j}(t,\phi)$  are used to interpolate the field during muon storage between the trolley runs.
The uncertainty on the interpolation is estimated from both the $k$ and $k+1$ maps and a Brownian bridge random walk model.
The procedure produces  interpolated storage volume field maps $\omega_{p}'(x,y,\phi)$ in terms of the equivalent shielded proton frequency throughout the \runone data-taking periods.

\begin{figure}[h]
\centering
\includegraphics[width=\columnwidth]{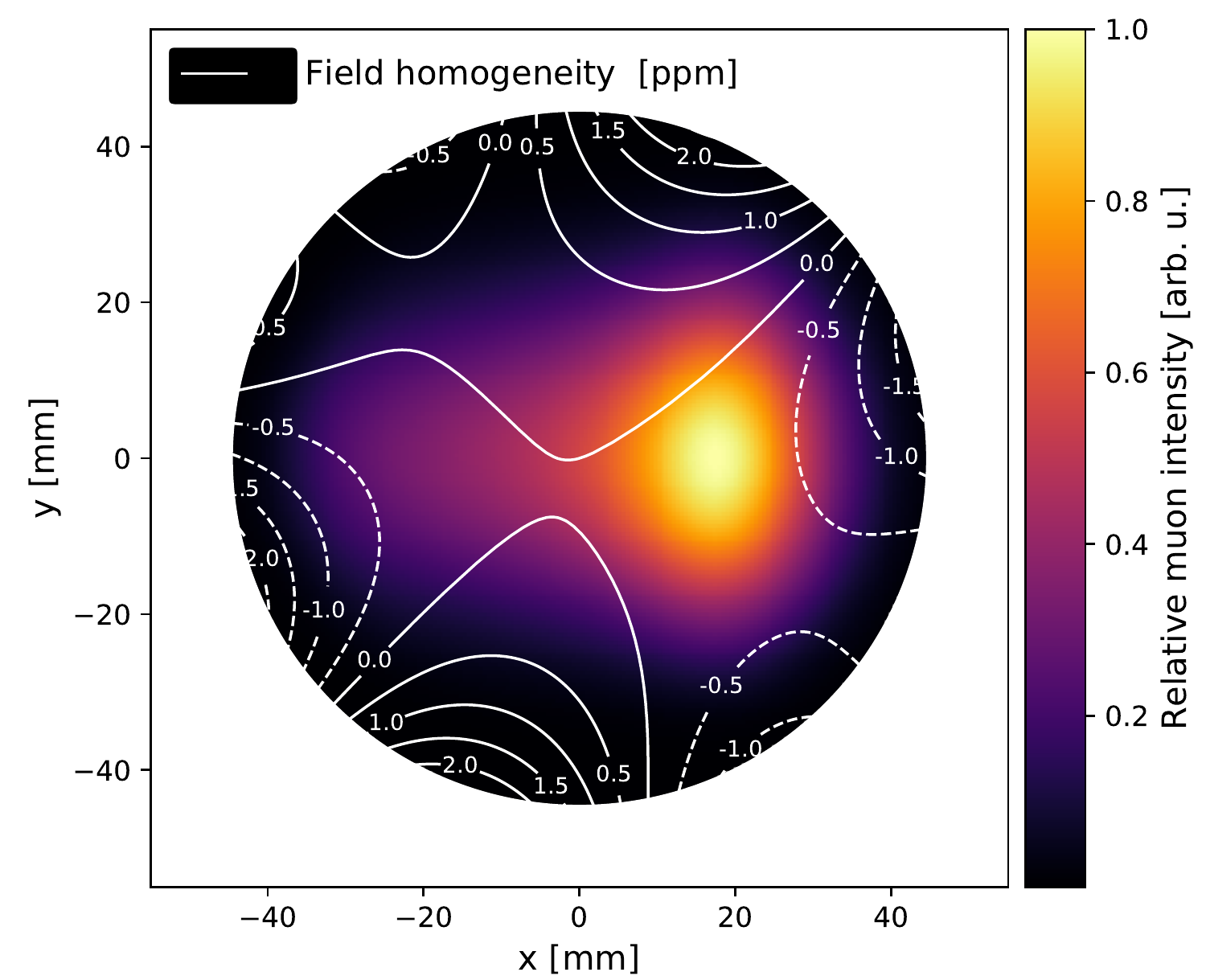}
\caption{Azimuthally averaged magnetic field contours $\omega_{p}'(x,y)$ overlaid on the time and azimuthally averaged muon distribution $M(x,y)$.}\label{fig:field}
\end{figure}

{\bf Muon weighting} ($M(x,y,\phi)$): Averaging of the magnetic field  weighted by the muon distribution in time and space uses the detected positron rates and the muon beam distribution measured by the trackers. The interpolated field maps are averaged over periods of roughly 10\,s and weighted by the number of detected positrons during the same period.
The SR guide fields introduce azimuthal dependencies of the muon distribution $M(x,y,\phi)$.
We determine the muon-weighted average magnetic field by summing the field moments $m_i$ multiplied by the beam-weighted projections $k_i$ for every three-hour interval over which the tracker maps and field maps are averaged. Along $y$, the beam is highly symmetric and centered, and the skew field moments (derivatives with respect to $y$) are relatively small. The azimuthally averaged centroid of the beam is displaced radially,  leading to relative weights for the field dipole, normal quadrupole, and normal sextupole of $k_i = 1.0, 0.15$, and $0.09$, respectively. An overlay of the azimuthally averaged field contours on the muon distribution is shown in Fig.~\ref{fig:field}. The combined total uncertainty of $\tilde\omega_p^\prime$  from probe calibrations, field maps, tracker alignment and acceptance, calorimeter acceptance, and beam dynamics modeling is 56\,ppb.

$\bm{B_{k}}$ and $\bm{B_{q}}$:
Two fast transients induced by the dynamics of charging the ESQ system and firing the SR kicker magnet slightly influence the actual average field seen by the beam compared to its NMR-measured value as described above and in Ref.~\cite{\field}.  An eddy current   induced locally in the vacuum chamber structures by the kicker system produces a transient magnetic field in the storage volume.  A Faraday magnetometer installed between the kicker plates measured the rotation of polarized light in a terbium-gallium-garnet  crystal from the transient field to determine the correction $B_k$.

The second transient arises from charging the ESQs, where the Lorentz forces induce mechanical vibrations in the plates that generate magnetic perturbations.
The amplitudes and sign of the perturbations vary over the two sequences of eight distinct fills that occur in each 1.4\,s accelerator supercycle.
Customized  NMR probes measured these transient fields at several positions within one ESQ and at the center of each of the other  ESQs to determine the average field throughout the quadrupole volumes.
Weighting the  temporal behavior of the transient fields by the muon decay rate, and correcting for the azimuthal fractions of the ring coverage, 8.5\% and 43\% respectively, each transient provides final corrections $B_{k}$ and $B_{q}$ to \amu as  listed in Table~\ref{tb:results}.

\section{\label{sec:conclusion}Computing \amu\ and Conclusions}

\begin{table}
\begin{ruledtabular}
\begin{tabular}{lrrr}
Run & $\omega_a/2\pi$\,[Hz] & $\opprimetilde/2\pi$\,[Hz] & $\Rmuprime \times 1000$\\
\hline
1a & 229081.06(28) & 61791871.2(7.1) & 3.7073009(45)\\
1b & 229081.40(24) & 61791937.8(7.9) & 3.7073024(38)\\
1c & 229081.26(19) & 61791845.4(7.7) & 3.7073057(31)\\
1d & 229081.23(16) & 61792003.4(6.6) & 3.7072957(26)\\ \hline
Run-1 &  &  & 3.7073003(17)\\
\end{tabular}
\end{ruledtabular}
\caption{\runone group measurements of \wa, \opprimetilde, and their ratios \Rmuprime multiplied by 1000. See also Supplemental Material~\cite{suppl:mat:1}.}
\label{tb:Rmu}
\end{table}

\begin{figure}[h]
\centering
\includegraphics[width=\columnwidth]{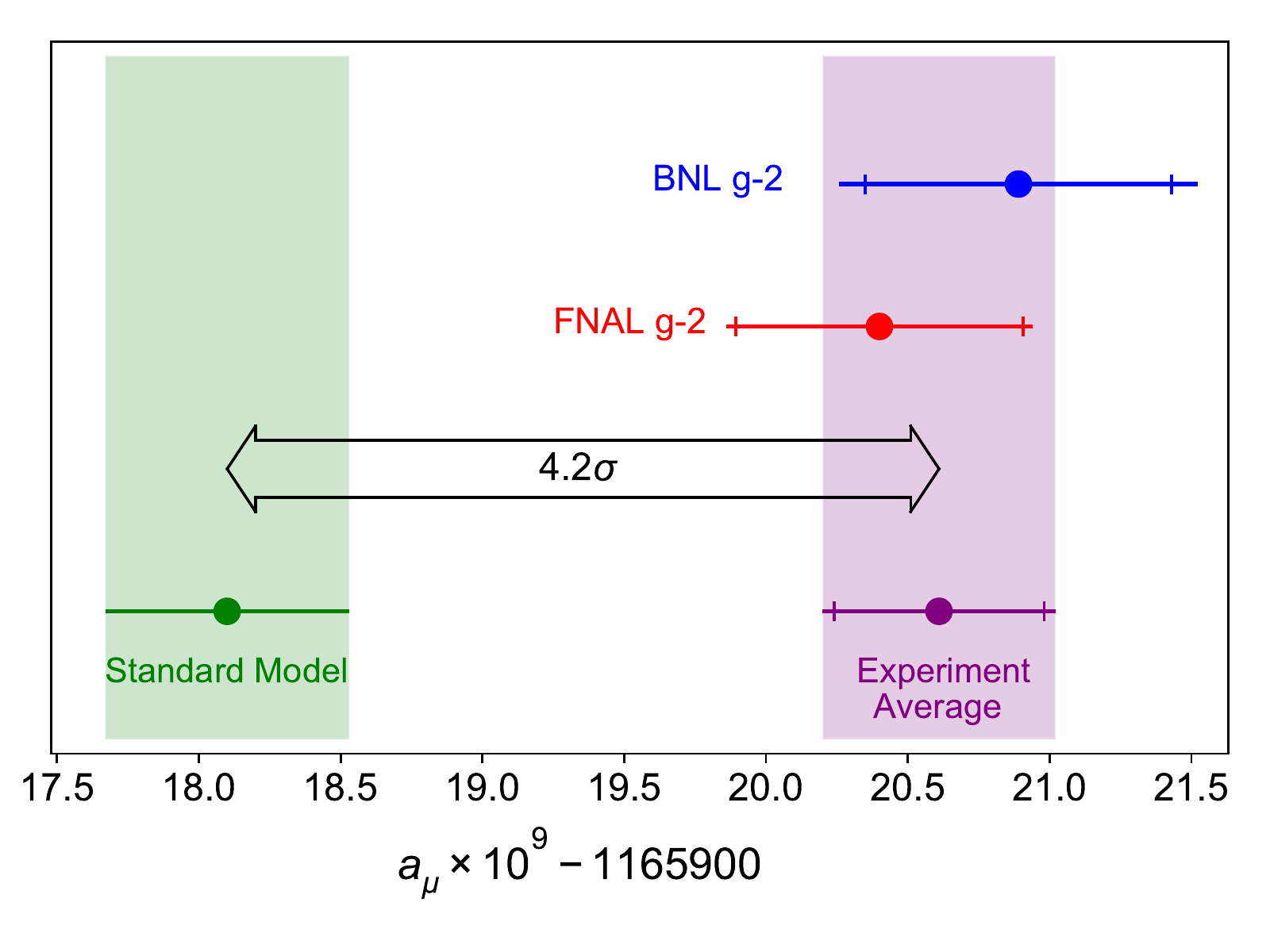}
\caption{From top to bottom:  experimental values of \amu from BNL E821, this measurement, and the combined average. The inner tick marks indicate the statistical contribution to the total uncertainties.  The Muon $g-2$ Theory Initiative recommended value~\cite{Aoyama:2020ynm} for the standard model is also shown.}\label{fig:results}
\end{figure}

Table \ref{tb:Rmu} lists the individual measurements of \wa and \opprimetilde, inclusive of all correction terms in Eq.~\ref{eq:Rcomponents},
for the four run groups, as well as their ratios, \Rmuprime (the latter multiplied by 1000).  The measurements are largely uncorrelated because the run-group uncertainties are dominated by the statistical uncertainty on \wa.  However, most systematic uncertainties for both  \wa\ and \opprimetilde measurements, and hence for the ratios \Rmuprime, are fully correlated across run groups.
The net computed uncertainties (and corrections) are listed in Table~\ref{tb:results}.
The fit of the four run-group results has a $\chi^2/\text{n.d.f.} = 6.8/3$, corresponding to $P(\chi^2) = 7.8\%$;
we consider the $P(\chi^2)$ to be a plausible
 statistical outcome
and not indicative of incorrectly estimated uncertainties.
The weighted-average value is
\Rmuprime = 0.0037073003(16)(6), where the first error is statistical and the second is systematic~\footnote{The E821 results for the field measurements were expressed in terms of the equivalent free proton precession frequency, resulting in $R^{\text{free}}_\mu(\text{BNL}) = 0.0037072063(20)$.
Expressing the field instead in terms of the proton shielded in water at $34.7^{\circ}$\,C results in  \Rmuprime(BNL) = 0.0037073019(20).}.
From Eq.~\ref{eq:amueq}, we arrive at a determination of the muon anomaly
\begin{equation*}
a_\mu({\rm FNAL}) = 116\,592\,040(54) \times  10^{-11}  ~~~ (\text{0.46\,ppm}),
\end{equation*}
where the statistical, systematic, and fundamental constant uncertainties that are listed in Table~\ref{tb:results} are combined in quadrature.
Our result differs from the SM value by $3.3\,\sigma$ and agrees with the BNL E821 result.  The combined experimental (Exp) average\footnote{We have carefully assessed any and all possible correlations to E821 at BNL and have concluded there are no important correlations that would impact a weighted average to obtain a correct combined result.There are also no non-negligible correlations between $\amu(\text{Exp})$ and $\amu(\text{SM})$.} is
\begin{equation*}
\amu(\text{Exp}) = 116\,592\,061(41) \times 10^{-11}   ~~~(0.35\,\text{ppm}).
\end{equation*}
The difference, $\amu(\text{Exp}) - \amu(\text{SM})= (251 \pm 59) \times 10^{-11}$, has a significance of  $4.2\,\sigma$.
These results are displayed in Fig.~\ref{fig:results}.

\begin{table} [h]
\begin{ruledtabular}
\begin{tabular}{lrr}
Quantity & Correction terms & Uncertainty\\
 &(ppb) & (ppb)\\
\hline
\wam (statistical) & -- & 434\\
\wam (systematic) & -- & 56\\
\hline
$C_e$ & 489 & 53\\
$C_p$ & 180 & 13\\
$C_{ml}$ & -11 & 5\\
$C_{pa}$ & -158 & 75\\
\hline
$f_{\text{calib}}\langle \omega_{p}’(x,y,\phi)\times M(x,y,\phi)\rangle$ & -- & 56\\
$B_k$ & -27 & 37\\
$B_q$ & -17 & 92\\ \midrule
$\mu'_p(34.7^\circ)/\mu_e$  & -- & 10\\
$m_\mu/m_e$  & -- & 22\\
$g_e/2$ & -- & 0\\
\hline
Total systematic & -- & 157 \\
Total fundamental factors & -- & 25 \\
Totals &544 & 462\\
\end{tabular}
\end{ruledtabular}
\caption{Values and uncertainties of the \Rmuprime correction terms in Eq.~\ref{eq:Rcomponents}, and uncertainties due to the constants in Eq.~\ref{eq:amueq} for \amu. Positive $C_i$  increase \amu and positive $B_i$ decrease \amu.}
\label{tb:results}
\end{table}

In summary, the findings here confirm the BNL experimental result and the corresponding experimental average increases the significance of the discrepancy between the measured and SM predicted \amu to $4.2\,\sigma$.  This result will further motivate the development of SM extensions, including those having new couplings to leptons.

Following the \runone measurements, improvements to the temperature in the experimental hall have led to greater magnetic field and detector gain stability. An upgrade to the kicker enables the incoming beam to be stored in the center of the storage aperture, thus reducing various beam dynamics effects. These changes, amongst others, will lead to higher precision in future publications.

\section{Acknowledgments}

We thank the Fermilab management and staff for their strong support of this experiment, as well as
the tremendous support from our university and national laboratory engineers, technicians, and workshops.
We are indebted to Akira Yamamoto, Lou Snydstrup and Chien Pai who provided
critical advice and engineering about the storage ring magnet and helped shepherd its transfer from Brookhaven to Fermilab.
Greg Bock and Joe Lykken set the blinding clock and diligently monitored its stability.
This result could not be interpreted without the worldwide theoretical effort
to establish the standard model prediction, and in particular the recent work by the
Muon \gmtwo Theory Initiative.

The Muon \gmtwo Experiment was performed at the Fermi National
Accelerator Laboratory, a U.S. Department of Energy, Office of
Science, HEP User Facility. Fermilab is managed by Fermi Research
Alliance, LLC (FRA), acting under Contract No. DE-AC02-07CH11359.
Additional support for the experiment was provided by the Department
of Energy offices of HEP and NP (USA), the National Science Foundation
(USA), the Istituto Nazionale di Fisica Nucleare (Italy), the Science
and Technology Facilities Council (UK), the Royal Society (UK), the
European Union's Horizon 2020 research and innovation program under
the Marie Sk\l{}odowska-Curie Grant Agreements No. 690835,
No. 734303, the National Natural Science Foundation of China
(Grant No. 11975153, 12075151), MSIP, NRF and IBS-R017-D1 (Republic of Korea),
and the German Research Foundation (DFG) through the Cluster of
Excellence PRISMA+ (EXC 2118/1, Project ID 39083149).

%

\end{document}